\pdfoutput=1

\documentclass[11pt]{article}

\usepackage[final]{acl}

\usepackage{times}
\usepackage{latexsym}
\usepackage{amsmath}
\usepackage{algorithm}
\usepackage{algorithmic}
\usepackage{multicol}
\usepackage{multirow}
\usepackage{colortbl}
\usepackage{float}
\usepackage{booktabs}
\usepackage{tabularx}
\usepackage{diagbox}

\usepackage[T1]{fontenc}

\usepackage[utf8]{inputenc}

\usepackage{microtype}

\usepackage{inconsolata}

\usepackage{graphicx}
\usepackage{marvosym}

\makeatletter
\def\thanks#1{\protected@xdef\@thanks{\@thanks
        \protect\footnotetext{#1}}}
\makeatother

\title{IDEAW: Robust Neural Audio Watermarking with \\Invertible Dual-Embedding}
\makeatletter\def\Hy@Warning#1{}\makeatother 
\author{Pengcheng Li\textsuperscript{\rm 1,2\ddag}\thanks{{\ddag} \  Equal contribution.}, Xulong Zhang\textsuperscript{\rm 1\ddag}, Jing Xiao\textsuperscript{\rm 1}, Jianzong Wang\textsuperscript{\rm 1\textsuperscript{\Letter}}\thanks{\textsuperscript{\scriptsize{\Letter}} Corresponding author.} 
\\
\textsuperscript{\rm 1}\textit{Ping An Technology (Shenzhen) Co., Ltd.} \\
    \textsuperscript{\rm 2}\textit{University of Science and Technology of China}\\
    \texttt{lipengcheng@ustc.edu, zhangxulong@ieee.org,} \\ \texttt{xiaojing661@pingan.com.cn, jzwang@188.com}
    }

\begin{document}
\maketitle
\begin{abstract}
The audio watermarking technique embeds messages into audio and accurately extracts messages from the watermarked audio. Traditional methods develop algorithms based on expert experience to embed watermarks into the time-domain or transform-domain of signals. With the development of deep neural networks, deep learning-based neural audio watermarking has emerged. Compared to traditional algorithms, neural audio watermarking achieves better robustness by considering various attacks during training. However, current neural watermarking methods suffer from low capacity and unsatisfactory imperceptibility. Additionally, the issue of watermark locating, which is extremely important and even more pronounced in neural audio watermarking, has not been adequately studied. In this paper, we design a dual-embedding watermarking model for efficient locating. We also consider the impact of the attack layer on the invertible neural network in robustness training, improving the model to enhance both its reasonableness and stability. Experiments show that the proposed model, \textbf{IDEAW}, can withstand various attacks with higher capacity and more efficient locating ability compared to existing methods. The code is available at \url{https://github.com/PecholaL/IDEAW}.
\end{abstract}

\section{Introduction}

\begin{figure}[htb]
    \centering
    \includegraphics[width=0.48\textwidth]{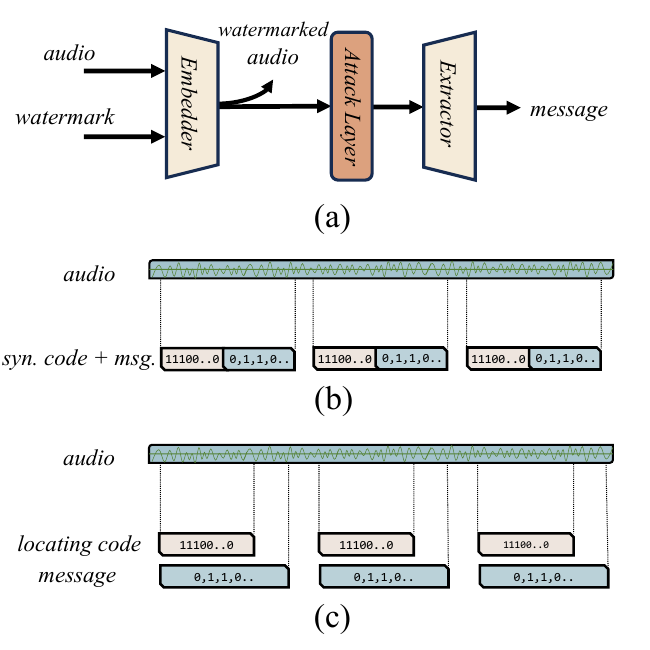}
    \caption{(a) Pipeline of robust neural audio watermarking. (b) Embedding strategy of existing methods. (c) Dual-embedding strategy of IDEAW.}
    \label{fig:intro}
\end{figure}

Digital watermarking \cite{singh2023classical} embeds messages indicating the ownership or authenticity into multimedia like images, video and audio, imperceptibly. This technique is widely used for ownership statements and anti-counterfeit. Imperceptibility and robustness are the two most challenging requirements for digital watermarking, which means, it is expected to be hard to feel the presence of the embedded watermark with human perception, and the watermark can be preserved and extracted accurately even after the watermarked media has been subjected to unintentional damage or malicious removal attacks. 

Audio watermarking has been around for decades. Traditional techniques \cite{singha2022development,zhang2023m} embeds the watermark into either the time-domain or transform-domain of audio signals via algorithms designed based on expert knowledge \cite{prabha2022a}. The robustness of traditional watermarking methods generally stems from subjective design, which results in limitations. The advancement of deep learning brings new solutions to steganography \cite{hussain2020survey,chanchal2020comprehensive} and digital watermarking techniques \cite{amrit2022survey,singh2023comprehensive}. End-to-end neural watermarking model completes the embedding and extraction process in each training iteration and constrains the imperceptibility and the integrity of the extracted watermarks through the designed training objectives. The attack layer which simulates common damages on the watermarked media is introduced into the Embedder-Extractor (\textit{i.e.} encoder-decoder) structure to guarantee the robustness. 

Neural audio watermarking is currently in its early stages. As human auditory perception is sometimes more sensitive than visual perception \cite{lee2019surface}. It can also easily distinguish noise, making subtle alterations caused by watermarking to be perceived. As for the robustness of watermarking. The Embedding-Attacking-Extracting pipeline of neural audio watermarking is shown in Fig. \ref{fig:intro}(a), where the attack layer simulates various removal attacks on watermarked audio during training. Redundancy is required for embedding digital watermarks to enhance applicability. The same watermark is repeatedly embedded at various locations within an audio segment. However, this strategy raises the issue of locating. The embedding location of the watermark is unknown during extraction in practical scenarios.  Additionally, trimming and splicing cause changes in the watermarking location. Compared to the traditional method, the extraction of neural audio watermarking relies on the forward process of neural networks which induces a non-negligible time cost that increases with the complexity of the network. Existing methods typically use an exhaustive approach, extracting synchronization code and watermark message together step by step, as shown in Fig. \ref{fig:intro}(b). \textit{Localization efficiency} is an issue that neural audio watermarking must face. 

The symmetry of the embedding and extraction processes of watermarking provides the invertible neural network with ample opportunities, but the introduction of the attack layer disrupts the original symmetry \cite{liu2019a}. In other words, due to the presence of the attack layer, the output of the encoder (\textit{i.e.} watermarked audio) and the input of the decoder (\textit{i.e.} attack-performed watermarked audio) are inconsistent, while the encoder and decoder are opposing and share parameters, this mismatch limits the training effect.

In this paper, we propose a model called \textbf{I}nvertible \textbf{D}ual-\textbf{E}mbedding \textbf{A}udio \textbf{W}atermarking, \textbf{IDEAW}, which uses a dual-stage invertible neural network with a dual-embedding strategy to embed watermark message and synchronization code (referred to as \textit{locating code} in this paper) separately, as illustrated in Fig. \ref{fig:intro}(c). During extraction, we first extract the less computational cost locating code. Upon successful matching, the extraction of the message which has more computational cost is conducted. This also makes it possible to enlarge the capacity of watermarking flexibly. To alleviate the asymmetric impact caused by the attack layer, we apply a balance block to enhance the training stability while preserving the characteristics of the invertible neural network. Our contributions in this paper can be summarized as follows:
\begin{itemize}
    \item Considering the characteristics of neural audio watermarking, we design a dual-embedding strategy to embed the watermark message and locating code separately, to accelerate the locating process. 
    \item We introduce the balance block to alleviate the asymmetry caused by the attack layer between the invertible network and retain the symmetry of the invertible neural network. 
    \item The proposed watermarking model is able to embed more bits of watermark while ensuring imperceptibility and robustness. 
\end{itemize}


\begin{figure*}[ht]
    \centering
    \includegraphics[width=1\textwidth]{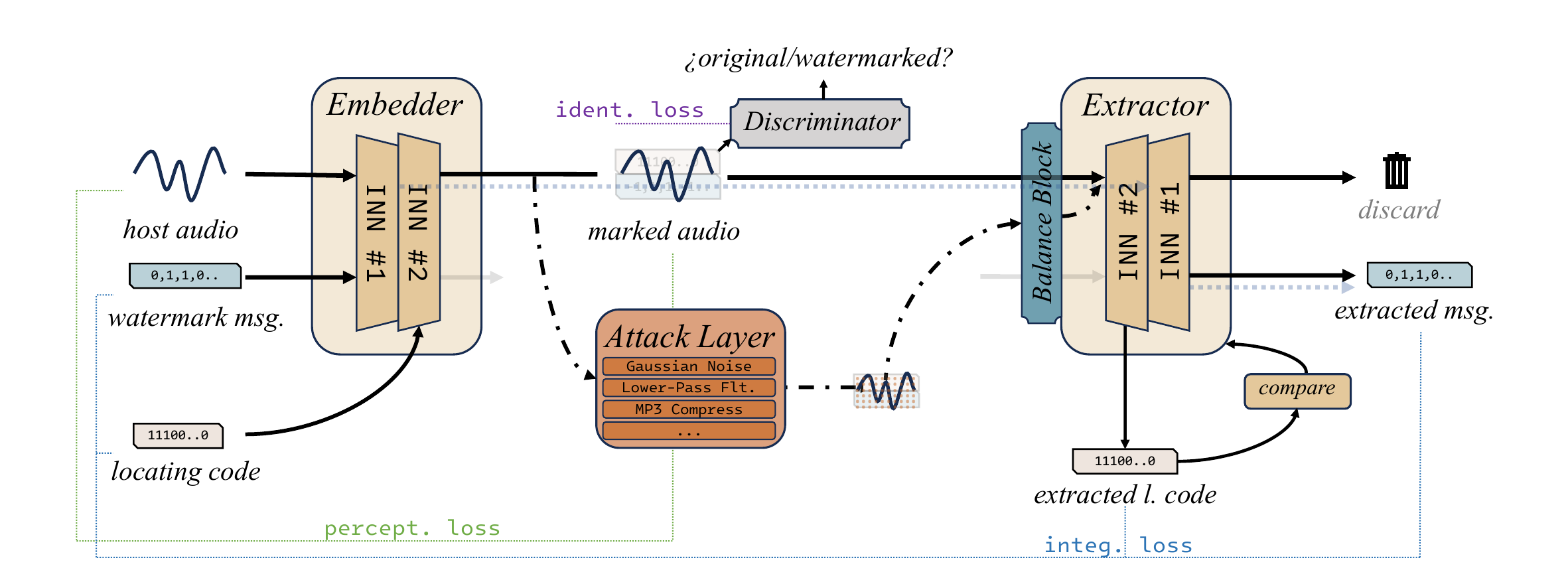}
    \caption{Architecture of IDEAW and the training objectives.}
    \label{fig:arch}
\end{figure*}

\section{Related Work}
\subsection{Neural Audio Watermarking}
The neural audio watermarking model is typically composed of two neural networks for watermark embedding and extraction in the Short Time Fourier Transform (STFT) or Discrete Wavelet Transform (DWT) domain. Pavlovi\'c \textit{et al.} \cite{pavlovic2020speech} design an Encoder-Decoder architecture for speech watermarking, where the encoder and the decoder form an adversarial relationship and are trained together. Hereafter, they introduce an attack layer to their previous work to form an Encoder-Attack Layer-Decoder structure, enhancing the robustness of the DNN-based speech watermarking \cite{pavlovic2022robust}. WavMark \cite{chen2023wavmark} considers the extraction as the inverse process of embedding the watermark, and leverages an invertible neural network to perform embedding and extraction for audio watermarking. DeAR \cite{liu2023dear} focuses on the threat of re-recording attack to audio watermarking, modelling re-recording as several differentiable processes. In addition to imperceptibility and robustness, capacity and locating effectiveness are also important criteria for evaluating audio watermarking methods. Compared to traditional watermarking methods, most neural audio watermarking methods suffer low capacity and do not consider the high consumption resulting from the locating process.

\subsection{Invertible Neural Network}
NICE \cite{dinh2015nice} firstly introduces the conception of an Invertible Neural Network (INN), a normalizing flow-based framework, learning a transformation that converts data that follows the original distribution to a predefined distribution. Dinh \textit{et al.} improve the performance of INN through convolutional layers in Real NVP \cite{dinh2017density}. Ardizzone \textit{et al.} introduce the conditional INN (cINN) \cite{ardizzone2019guided} to establish control over the generation. Behrmann \textit{et al.} \cite{behrmann2019invertible} utilize ResNet as the Euler discretization of ordinary differential equations and prove that the invertible ResNet can be constructed by changing the normalization mechanism. INN has been widely applied in generation \cite{dinh2015nice,dinh2017density,kingma2018glow,ouderaa2019reversible}, image super-resolution \cite{lugmayr2020srflow}, image compression \cite{wang2020modeling}, image-to-image translation \cite{ouderaa2019reversible}, digital steganography \cite{lu2021large,mou2023large}, \textit{etc}.

\section{Method}
\subsection{Overall Architecture of IDEAW}
Fig. \ref{fig:arch} showcases the architecture of our proposed audio watermarking model, IDEAW, which follows an \textit{Embedder-Attack Layer-Extractor} structure, where the embedder and extractor are carefully designed as dual-stage structures for embedding messages and locating codes at vertical latitude separately. In the embedding process, an $L_m-bit$ binary watermark message $m\in\{0,1\}^{L_m}$ is embedded into a fixed-length audio chunk $x$ in the STFT domain via the first stage INN, then the second stage INN embeds a binary locating code $c\in\{0,1\}^{L_c}$ into the audio containing $m$ from the former step. The final watermarked audio is reconstructed through Inverse Short-Time Fourier Transform (ISTFT). A discriminator distinguishes between the host audio and the watermarked audio to guarantee the imperceptibility of watermarking. In the extraction process, the two-stage extractor first extracts $c$ from the watermarked audio and then $m$, in the reverse order of embedding. An attack layer is introduced to enhance the robustness to various removal attacks. The extractor must accurately extract $c$ and $m$ from the attack-performed watermarked audio. The balance block aims to alleviate the asymmetry introduced by the attack layer, preserving the symmetry of INN.

\subsection{Dual-stage INN for Dual-Embedding}
\begin{figure}[tb]
    \centering
    \includegraphics[width=0.48\textwidth]{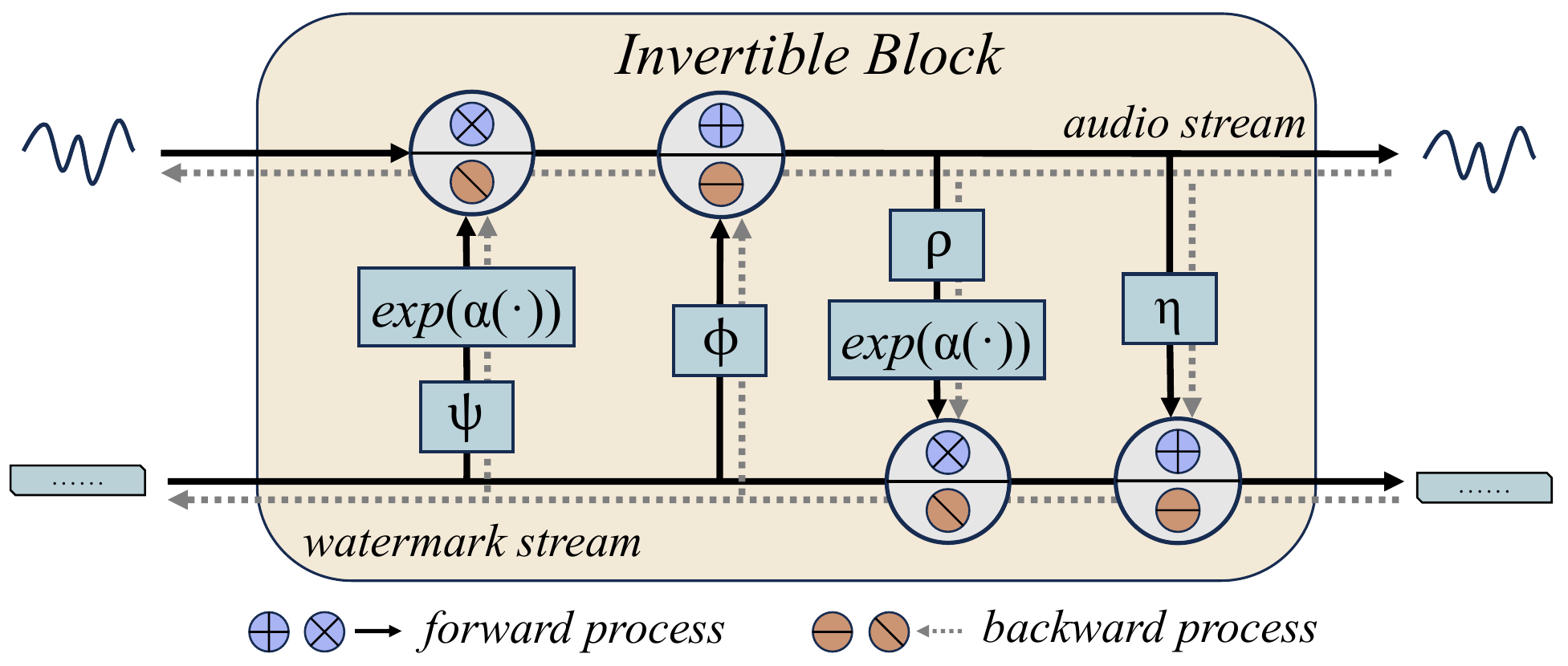}
    \caption{Structure and forward/backward processes of the invertible block.}
    \label{fig:inn}
\end{figure}

To vertically separate the locating code and message, a dual-stage INN is designed, with each stage designated as \texttt{INN$_{\#1}$} for the embedding and extraction of watermark message, and \texttt{INN$_{\#2}$} for that of locating code. Each INN consists of several invertible blocks that take transformed audio and the watermark (\textit{i.e.} the watermark message or locating code) as inputs, producing two outputs. We refer to these input-output pairs as two data streams: the audio stream and the watermark stream. The audio stream outputs watermarked audio during the embedding process, while the watermark stream provides a watermark message during extraction. Fig. \ref{fig:inn} showcases the architecture of the invertible block, in which the block processes the data as illustrated in Eq. \ref{eq:inn}:
\begin{equation}
\label{eq:inn}
\begin{split}
& x^{i+1}=x^{i}\odot exp(\alpha(\psi(s^{i})))+\phi(s^{i}) \\
& s^{i+1}=s^{i}\odot exp(\alpha(\rho(x^{i+1})))+\eta(x^{i+1})
\end{split}
\end{equation}
where $x$ denotes the host data including the original audio fed to \texttt{INN$_{\#1}$} and the audio with message embedded fed to \texttt{INN$_{\#2}$}, $s$ denotes the secret data including the message fed to \texttt{INN$_{\#1}$} and the locating code fed to \texttt{INN$_{\#2}$}. $\alpha(\cdot)$ is a sigmoid function, while $\psi(\cdot)$, $\phi(\cdot)$, $\rho(\cdot)$ and $\eta(\cdot)$ are subnets which are constructed from dense blocks.

In the first embedding stage, \texttt{INN$_{\#1}$} embeds the watermark message into the host audio, where the audio and bit sequences are transformed into the time-frequency domain via STFT at first, respectively. Then \texttt{INN$_{\#2}$} embeds the locating code into the audio stream output of \texttt{INN$_{\#1}$}. The dual-embedding can be described by Eq. \ref{eq:embed}. Subscripts $_{a}$ and $_{wm}$ of $\texttt{INN}(\cdot)$ denote the output of the audio stream and watermark stream of INN, respectively.
\begin{equation}
\label{eq:embed}
x_{wmd}=\texttt{INN}_{\#2}(\texttt{INN}_{\#1}(x,m)_a,c)_a
\end{equation}

The extraction process of IDEAW is exactly the opposite of the embedding process. \texttt{INN$_{\#2}$} firstly extract the locating code from the watermarked audio, then the output of the audio stream from \texttt{INN$_{\#2}$} is sent to \texttt{INN$_{\#1}$} to extract watermark message. Eq. \ref{eq:extract} describes the dual extraction to obtain the embedded message.
\begin{equation}
\label{eq:extract}
\hat{m}=\texttt{INN}_{\#1R}(\texttt{INN}_{\#2R}(x_{wmd},x_{aux_2})_a,x_{aux_1})_{wm}
\end{equation}
where $x_{aux}$ represents the randomly sampled signal which is fed to the message stream of INN. The subscript \textit{R} represents the reverse process of the reversible network as the extraction process employs the same network and parameters as the embedding process, only with the order reversed.

During training, the message is extracted by $\texttt{INN}_{\#1R}$ after each stage of embedding, then compared with the original message $m$. The locating code is extracted from the final watermarked audio via $\texttt{INN}_{\#2R}$ and compared with the original locating code. The integrity loss is as follows.
\begin{equation}
\label{eq:integrity}
\begin{split}
&\mathcal{L}_{integ}=||\hat{m}-m||_2 \\
&\thickspace\thickspace\thickspace
+||\texttt{INN}_{\#1R}(\texttt{INN}_{\#1}(x,m)_a,x_{aux_1})_{wm}-m||_2 \\
&\thickspace\thickspace\thickspace
+||\texttt{INN}_{\#2R}(x_{wmd},x_{aux_2})_{wm}-c||_2
\end{split}
\end{equation}

In practical scenarios, watermarks are embedded into several segments of audio as shown in Fig. \ref{fig:intro} (c). However, watermarked audio may be trimmed or spliced (known as the de-synchronization attacks \cite{mushgil2018efficient}), making it challenging to determine the watermark location. The extractor extracts and matches the locating code quickly as \texttt{INN$_{\#2}$} is lighter and costs less computation than \texttt{INN$_{\#1}$}, extracting the watermark message only if the locating code is matched. In addition, a similar training manner to the shift module in WavMark \cite{chen2023wavmark} is deployed in our proposed model, which helps the extractor to gain the ability to extract watermark from a proximity location. 

\subsection{Imperceptibility Guaranty}
As one of the most important indicators for evaluating digital watermarking, imperceptibility ensures that the embedding of the watermark cannot be distinguished from human auditory perception. In IDEAW, we take two measures to ensure and improve the imperceptibility of watermarking. The perceptual loss intuitively requires that the difference between the watermarked audio and the original host audio is as narrow as possible, that is, reducing the impact of watermarking on the host audio. The perceptual loss is shown in Eq. \ref{eq:percept}. 
\begin{equation}
\label{eq:percept}
\mathcal{L}_{percept}=||x_{wmd}-x||_2
\end{equation}

What's more, we leverage a discriminator to distinguish the watermarked audio from the origin audio. Essentially, the discriminator is a binary classifier which classifies host and watermarked audio labeled with $0$ and $1$ (denoted as label $y$) respectively. The discriminator aims to identify the watermarked audio while the embedder tries to hide the watermark so that the embedder and the discriminator form an adversarial relationship and mutually force each other during training \cite{goodfellow2020generative}. The discriminate loss and identify loss is as follows:
\begin{equation}
\label{eq:discr}
\mathcal{L}_{discr}=-y\cdot log(D(x))-(1-y)\cdot log(1-D(x))
\end{equation}

\begin{equation}
\label{eq:ident}
\mathcal{L}_{ident}=-log(1-D(x_{wmd}))
\end{equation}

\subsection{The Robustness and Symmetry of INN}
The robustness of a watermarking method is critical to the attacker's ability to effectively remove the embedded watermarks. Generally, attackers try to remove the watermark from the watermarked audio through several removal techniques including passing the audio through filters only to maintain the information that can be perceived by humans, or adding noises to the watermarked audio to interfere with accurate extraction. Effective watermark removal requires that the watermark cannot be extracted correctly from the watermark-removed audio and that the watermark-removed audio should be usable, \textit{i.e.} the listening quality should not significantly decrease, otherwise the removal of the watermark is meaningless.

In order to endow the neural audio watermarking with robustness against various watermark removal attacks, an attack layer is incorporated into the watermarking model and trained alongside the embedding and extraction networks. The attack layer subjects the watermarked audio (\textit{i.e.} the output from the embedder) to a variety of attacks. Subsequently, the extractor attempts to extract the locating code and watermark message from the audio that has undergone attack as accurately as possible. Considering the effectiveness of the watermark removal, the predefined attacks should ensure the quality of attack-performed watermarked audio does not degrade too much. These attacks include Gaussian additive noise, lower-pass filter, MP3 compression, quantization, resampling, random dropout, amplitude modification and time stretch. 

However, the introduction of the attack layer disrupts the symmetry of the entire embedding-extraction process, impacting the training of INN. We define the integral dual-embedding process and extraction process as $f(\cdot)$ and $f^{-1}(\cdot)$ respectively. During the embedding process, IDEAW performs $f(x,m,c)=w$, where $x,m,c,w$ are host audio, watermark message, locating code and watermarked audio as mentioned above. And $w$ follows the distribution of watermarked audio $P_W(w)$ \cite{ma2022toward}. While in the extraction process, thanks to the parameter-sharing between the embedder and the extractor, $m$ and $c$ can be easily sampled with $m,c=f^{-1}(w)$ according to $P_W(w)$. But the including of the attack layer causes changes in $P_W(w)$ leading to a $P_{W'}(w')$, while the extraction process is still based on the unchanged $P_W(w)$, which affects the performance of watermarking. The parameter-sharing strategy of INN limits the extractor's learning ability to adapt to the attacked audio, resulting in distorted watermarking performance.

To simultaneously maintain the parameter-sharing of INN and the symmetry of INN's training, a balance block is employed to mitigate the asymmetry caused by the attack layer and stabilize the model's symmetric structure. The balance block consists of a group of dense blocks \cite{huang2017densely} which process the input to equal-size output, providing extra trainable parameters for the extractor. This manner learns to transform the attack-performed watermarked audio distribution $P_{W'}(w')$ to the revised distribution $P_{\hat{W}(\hat{w})}$ that close to the expected distribution $P_W(w)$ under the guidance of the mentioned integrity loss, to counteract the effects of the offset introduced by the attack layer without bothering the embedder.

\subsection{Training Strategy}
Simultaneously ensuring accuracy, imperceptibility and robustness is sticky. Therefore, the training of IDEAW is divided into two stages:
 \begin{itemize}
    \item The first stage only considers the imperceptibility and the watermark integrity of extraction, aiming to build a dual-stage INN that can embed the watermark imperceptibly and extract the watermark accurately. 
    \item In the second stage, the requirement for the robustness of watermarking is introduced. The attack layer and balance block are incorporated into the model, and the entire model is trained collectively. 
 \end{itemize}

The same total loss function, as introduced in Eq. \ref{eq:loss}, is applied for both training stages, where $\lambda_1$, $\lambda_2$ and $\lambda_3$ are weights of each component, but note that the second stage contains more trainable parameters. The discriminator is trained along with the watermarking model in each iteration with the loss function in Eq. \ref{eq:discr}. The pseudo script of the training process as well as the acquisition of the total loss is shown in Section \ref{appendix:alg} of the appendix. 
\begin{equation}
\label{eq:loss}
\mathcal{L}_{total} = \lambda_1\mathcal{L}_{integ} + \lambda_2\mathcal{L}_{percept} + \lambda_3\mathcal{L}_{ident}
\end{equation}

\begin{table}[b]
    \centering
    \begin{tabular}{c c c c}
        \toprule
        \multirow{2}{*}{ Method }
        & \footnotesize{$SNR\uparrow$} 
        & \footnotesize{$ACC\uparrow$}
        & \footnotesize{$Capacity\uparrow$} \\
        & \scriptsize{$(dB)$} & \scriptsize{$(\%)$} & \scriptsize{$(bps)$} \\
        \midrule
        DeAR
            & 26.18 & 99.61 & 8.8 \\
        \rowcolor{gray!15}
        \textbf{IDEAW}$_{10+10}$
            & 40.43 & 99.64 & 20 \\
        WavMark
            & 38.55 & 99.35 & 32 \\
        \rowcolor{gray!15}
        \textbf{IDEAW}$_{22+10}$
            & 37.72 & 99.52 & 32 \\
        \rowcolor{gray!25}
        \textbf{IDEAW$_{46+10}$}
            & 35.41 & 99.44 & 56 \\
        \bottomrule
    \end{tabular}
    \caption{Comparison of the basic metrics with baseline methods. The $ACC$ of IDEAW is calculated from the locating code and message. }
    \label{tab:basic}
\end{table}

\begin{table*}[htb]
\centering
    \begin{tabular}{cccccccccc}
    \toprule
    \multirow{2}{*}{\diagbox{Method}{Attack}} \quad 
    & GN
    & LF
    & CP
    & QZ
    & RD
    & RS
    & AM
    & TS
    \cr
    & \footnotesize{$35 dB$} 
    & \footnotesize{$5 kHz$}
    & \footnotesize{$64 kbps$}
    & \footnotesize{$2^9$}
    & \footnotesize{$0.1\%$}
    & \footnotesize{$200\%$}
    & \footnotesize{$90\%$}
    & \footnotesize{$90\%$}
    \cr
    \midrule
    DeAR
             & 99.61 & 99.04 & 99.55 
             & 99.63 & 99.61 & 99.62 
             & 99.61 & 99.31 
    \cr
    \rowcolor{gray!15}
    \textbf{IDEAW$_{10+10}$}
             & 99.47 & 98.62 & 99.41
             & 98.86 & 99.60 & 99.11 
             & 99.49 & 98.95 
    \cr
    WavMark
             & 97.84 & 98.54 & 98.81 
             & 96.60 & 98.68 & 98.42 
             & 99.29 & 95.35 
    \cr
    \rowcolor{gray!15}
    \textbf{IDEAW$_{22+10}$}
             & 99.33 & 98.53 & 99.15 
             & 98.73 & 99.23 & 99.02 
             & 99.48 & 98.82 
    \cr
    \rowcolor{gray!25}
    \textbf{IDEAW$_{46+10}$}
             & 98.72 & 98.12 & 99.00 
             & 98.61 & 98.84 & 98.87 
             & 99.42 & 98.66 
    \cr
    \bottomrule
    \end{tabular}
    \caption{Comparison of the robustness with baseline methods. The robustness is evaluated according to \textbf{ACC(\%)}$\uparrow$ under different watermark removal attacks. }
    \label{tab:robustness}
\end{table*}

\section{Experiment}
\subsection{Settings}
\subsubsection{Dataset and Implementation}
IDEAW is trained on VCTK corpus \cite{vctk2016} and FMA corpus \cite{defferrard2017fma}. VCTK comprises over 100 hours of multi-speaker speech data, while FMA contains a large amount of music audio. These two types of data are prevalent in scenarios where audio watermarking is commonly applied. All the audio is resampled to 16,000 Hz and split into 1-second segments during training. STFT and ISTFT operations with parameters of $\{n\_fft=1000,hop\_length=250,win\_length=1000\}$ perform the inter-domain transformation.

Two Adam optimizers \cite{kingma2015adam} with $\{\beta_1=0.9, \beta_2=0.99, \epsilon=10^{-8}, lr=10^{-5}\}$ (with a StepLR scheduler) are leveraged for the introduced two-stage training. Super-parameters $\lambda_1$, $\lambda_2$ and $\lambda_3$ in total loss (Eq. \ref{eq:loss}) are set to $1$, $0.1$ and $0.1$, respectively. Each stage contains 100,000 iterations. The attacks are sample-wise, each audio segment in the batch undergoes different types of attacks during training. The configuration of the attack layer is shown in Section \ref{appendix:attack} of appendix. We set the length of the locating code to 10 bits, so there is about a $1/2^{10}$ probability of potential conflicts with extracted non-locating code bit sequences. In each batch, the locating code and watermark message are randomly generated to ensure that the trained model can handle any arrangement of 0-1 sequences. 

We select two existing neural audio watermarking works, WavMark \cite{chen2023wavmark} and DeAR \cite{liu2023dear}, as baselines.

\subsubsection{Metrics}
\textbf{Signal-to-noise ratio (SNR)} measures the impact of the watermarking. It is calculated as 10 times the logarithm of the ratio of host audio power to watermark noise power. \\
\textbf{Accuracy (ACC)} measures the differences between the extracted message and the ground-truth message. ACC is typically used to measure the robustness of watermarking methods. \\
\textbf{Capacity} is the number of watermark bits that can be embedded per second of audio while ensuring imperceptibility and ACC of the watermarking.

\subsection{Results}
\subsubsection{Overall Comparison}
The comparison of basic metrics of various watermarking methods is shown in Table \ref{tab:basic}. As each baseline watermarking model is designed for different capacities, and IDEAW owns the maximum one, we also train the other two IDEAW models which have similar or larger capacities as other methods to alleviate the impact of capacity on comparison. As the locating code and message are separated in our proposed model, we take the total length of the locating code and message as the capacity of IDEAW and the length of the locating code is fixed to 10 bits. We find that the watermarking model with lower capacity obtains better imperceptibility and higher accuracy. IDEAW gains considerable SNR and ACC when designed with a large payload.

\begin{figure}[htb]
    \centering
    \includegraphics[width=0.48\textwidth]{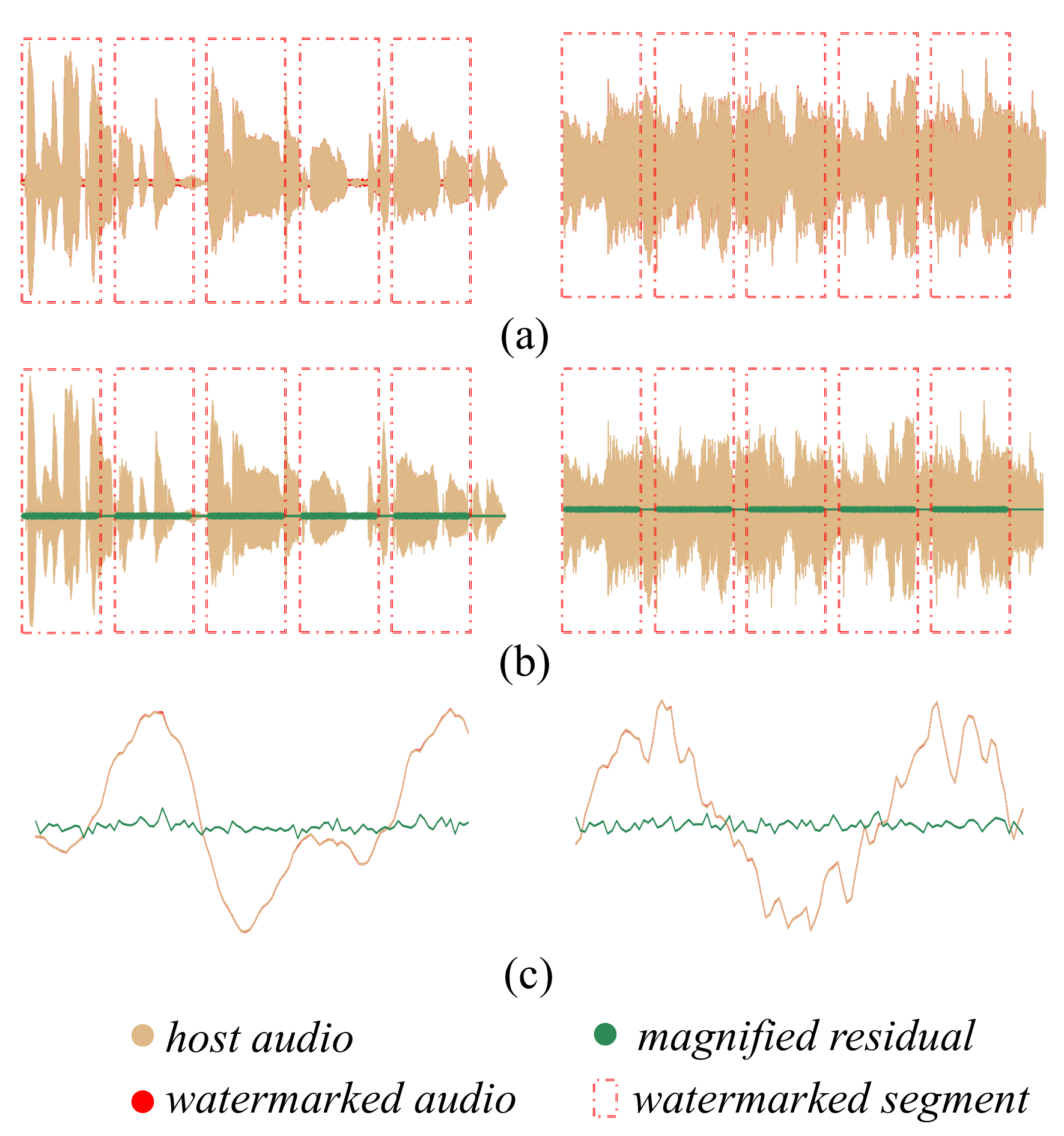}
    \caption{Waveforms of (a) host audio (foreground) and watermarked audio (background), (b) host audio and tenfold-magnified residual caused by watermarking, (c) local details (100 points) of the (a) and (b). The left audio is low-energy speech audio while the right is high-energy music audio.}
    \label{fig:exp}
\end{figure}

\begin{figure}[htb]
    \centering
    \includegraphics[width=0.48\textwidth]{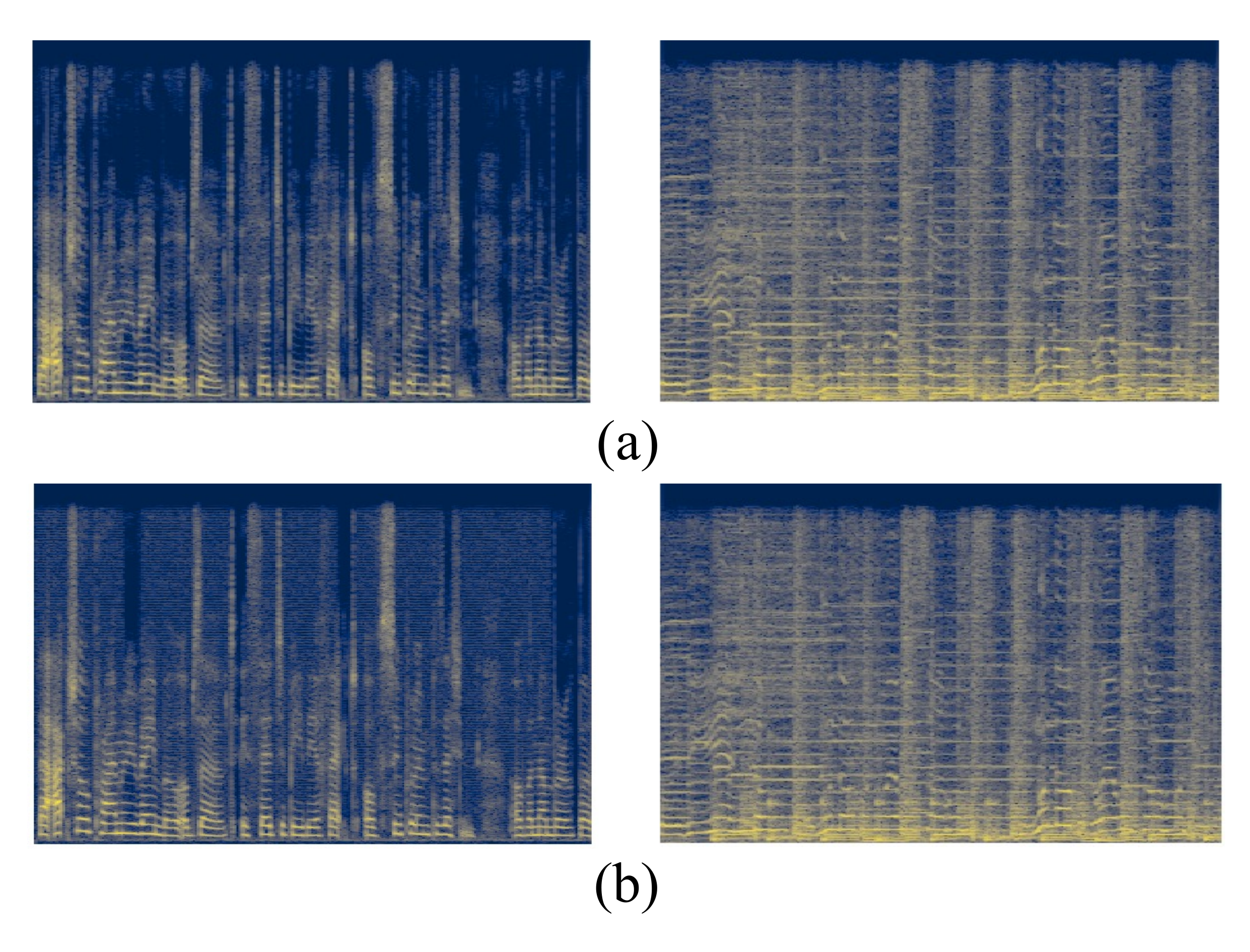}
    \caption{Linear-frequency power spectrograms of low-energy speech audio (left) and high-energy music audio (right). (a) the host audio, (b) the watermarked audio. }
    \label{fig:spect}
\end{figure}

Fig. \ref{fig:exp} illustrates the impact of the watermarking on a low-energy speech waveform and a high-energy music waveform. The same watermark is iteratively embedded into the audio. From Fig. \ref{fig:exp}(a), we can see that the original audio in the foreground and the watermarked audio in the background almost overlap. Fig. \ref{fig:spect} shows a comparison of the linear-frequency power spectrograms of the original and watermarked audio, illustrating the impact of the watermarking in the time-frequency domain. 

More watermarked audio samples, waveform samples and practical application examples are available at our demo page \url{https://largeaudiomodel.com/IDEAW}.

\subsubsection{Robustness Comparison}
We measure the watermark extraction accuracy of each model under different attacks to evaluate their robustness. Eight common attacks including Gaussian additive noise (GN), lower-pass filter (LF), MP3 compression (CP), quantization (QZ), random dropout (RD), resampling (RS), amplitude modification (AM) and time stretch (TS) are taken into consideration. As mentioned above,  watermark removal attacks need to consider the degree of damage to audio, and the strength settings of each attack continue the previous research settings \cite{chen2023wavmark}. The robustness evaluation results are shown in Table \ref{tab:robustness}. IDEAW shows comparable robustness to the baseline model while with a larger capacity. 

\begin{figure}[htb]
    \centering
    \includegraphics[width=0.48\textwidth]{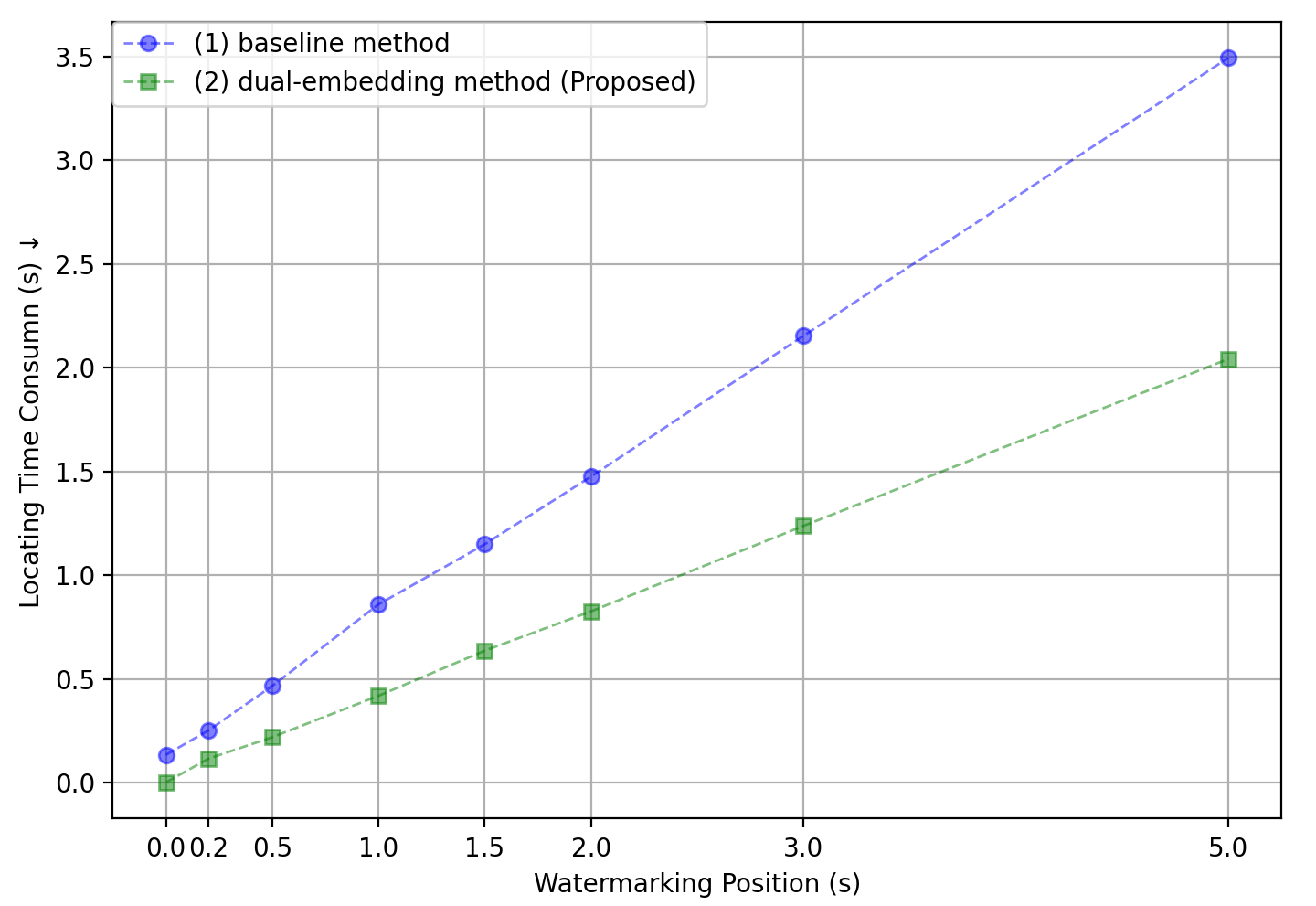}
    \caption{Comparison of locating time consumption for different methods at various watermark embedding locations (the location is indicated by the \textit{seconds} from the start of audio to the watermarking location).}
    \label{fig:locating}
\end{figure}

\begin{table}[b]
    \centering
    \begin{tabular}{c c c}
        \toprule
        \multirow{2}{*}{Method}
        & \quad \footnotesize{$SNR\uparrow$} \quad 
        & \quad \footnotesize{$ACC\uparrow$} \quad \cr
        & \quad \scriptsize{$(dB)$} & \ \scriptsize{$(\%)$}\cr
        \midrule
        \textbf{M1}
            & 32.30 & 99.40 \\
        \textbf{M2}
            & 35.27 & 98.70 \\
        \rowcolor{gray!25}
        \textbf{IDEAW}$_{46+10}$ 
            & 35.41 & 99.44 \\
        \bottomrule
    \end{tabular}
    \caption{Basic metrics comparison for ablation study.}
    \label{tab:ablation}
\end{table}

\begin{table*}[htb]
  \centering
    \begin{tabular}{cccccccccc}
    \toprule
    \multirow{2}{*}{\diagbox{Method}{Attack}} \quad 
    & GN
    & LF
    & CP
    & QZ
    & RD
    & RS
    & AM
    & TS
    \cr
    & \footnotesize{$35 dB$} 
    & \footnotesize{$5 kHz$}
    & \footnotesize{$64 kbps$}
    & \footnotesize{$2^9$}
    & \footnotesize{$0.1\%$}
    & \footnotesize{$200\%$}
    & \footnotesize{$90\%$}
    & \footnotesize{$90\%$}
    \cr
    \midrule
    \textbf{M1} 
             & 98.25 & 98.18 & 98.79 
             & 98.73 & 98.90 & 98.46 
             & 99.28 & 98.58 
    \cr
    \textbf{M2} 
             & 98.14 & 97.79 & 98.55 
             & 96.07 & 98.24 & 98.39 
             & 97.62 & 96.57 
    \cr
    \rowcolor{gray!25}
    \textbf{IDEAW$_{46+10}$}
             & 98.72 & 98.12 & 99.00 
             & 98.61 & 98.84 & 98.87 
             & 99.42 & 98.66 
    \cr
    \bottomrule
    \end{tabular}
    \caption{Comparison of the robustness in ablation study. The \textbf{ACC(\%)}$\uparrow$ of the extracted message and locating code under attack via each model.}
    \label{tab:ablationrobustness}
\end{table*}

\subsubsection{Locating Test}
To compare the time cost of each locating method, we embed the watermark (10-bit locating code or synchronization code and 46-bit message) at different locations and use different methods for locating. The tested methods include (1) the same as WavMark \cite{chen2023wavmark}, synchronization code and message are concatenated and embedded into the audio segment together via a single-stage INN model. The model is trained with the shift strategy. The synchronization code and message are extracted at the same time iteratively during locating. The step size is $10\%$ of the chunk size, the same as the baseline method. (2) the proposed method, only extracts the locating code during the locating process with a step size of $10\%$ of the chunk size. Method (1) builds a single stage INN watermarking model that has the same layer quantity as the proposed dual-stage model. Note that in order to eliminate the impact of errors during locating as we cannot guarantee the $ACC$ of the model with a carrying capacity of 56 bits in Method (1), we only perform extraction without further verification, until reaching the watermark location, as the former experiments show that each model can reach the ideal accuracy with designed capacity. 

The measurement of locating time for each method is conducted on the same device and takes an average of 100 processes for each watermarking location. The comparison of the time consumption of each method on the same device is shown in Fig. \ref{fig:locating}. The results show that the proposed method reduces time overhead by approximately $40\%\sim50\%$. Especially when the watermark is far from the head, the advantage of our dual-embedding strategy is more obvious.

\subsection{Ablation Study}
\subsubsection{Setting Up}
To validate the positive effects of the proposed balance block and the discriminator on IDEAW's performance, we conduct an ablation study. The following models related to the proposed methods are built. (1) \textbf{M1} (\textit{w/o discriminator}) removes the discriminator from the proposed method. (2) \textbf{M2} (\textit{w/o balance block}) removes the balance block from the proposed method during the robustness training. These models as well as the proposed model are trained on the same datasets and in the same manner. We measure the basic metrics and the robustness of each model. 

\subsubsection{Ablation Study Results}
The results of the ablation study are shown in Table \ref{tab:ablation} and Table \ref{tab:ablationrobustness}. The results show that the discriminator helps improve the quality and naturalness of the watermarked audio, as the model without discriminator has a degradation in the signal-to-noise ratio and comparable robustness to the proposed method. The balance block does alleviate the asymmetry caused by the attack layer in robustness training, enabling the model to gain better robustness and achieve higher accuracy. Overall, the introduction of the discriminator and balance block enhances the watermarking quality and the stability of the watermarking model. 

\section{Conclusion}
In this paper, we propose a neural audio watermarking model, IDEAW, which embeds the locating code and watermark message separately via a designed dual-stage INN. In response to the challenge of neural audio watermarking localization, the dual-embedding strategy avoids the huge computation of extracting synchronization code and message at the same time, accelerating the locating process. On the other hand, we make an effort to mitigate the asymmetry factor introduced by the attack layer. The balance block is leveraged to provide the extractor with a subtle training spatial while maintaining the advantage of the invertible network. Experimental results show that IDEAW achieves satisfactory performance from a comprehensive perspective of imperceptibility, capacity, robustness and locating efficiency.

\section*{Limitations}
This work focuses on the high overhead problem of localization of neural audio watermarks by innovatively separating the locating code from the watermark message and embedding/extracting them separately. However, we find that the dual-embedding watermarking method has the following limitations so far: 
\begin{itemize}
    \item The lengths of the locating code and messages of the trained model cannot be flexibly adjusted because the design of the dual-stage invertible neural network fixes the lengths of both, we cannot designate the bits starting at arbitrary lengths as synchronization codes like previous methods. 
    \item We find that the watermark embedded in low-energy audio is less imperceptible than in high-energy audio via the proposed model. The reason might be that, apart from selecting two types of datasets for training, the design of the model as well as its training strategy does not consider the carrier’s energy.
\end{itemize}

\section*{Future Works}
With the rapid development of social media and the increasing awareness of copyright, digital watermarking technology has promising prospects for widespread application. Neural audio watermarking remains a valuable area of research due to its robustness, independence from expert intervention, and scalability, compared to traditional methods. However, neural audio watermarking has not yet fully surpassed traditional digital watermarking methods, particularly in terms of capacity. Future works on neural audio watermarking may consider, but is not limited to, the following aspects: 
\begin{itemize}
    \item Maximizing capacity through novel neural network architectures or embedding strategies.
    \item Exploring adaptive approaches to narrow the performance gap between high-energy and low-energy audio.
    \item Considering implementing watermark embedding directly within the audio generation model to obtain generated audio with watermarks, rather than using post-processing watermarking methods.
\end{itemize}

\section*{Acknowledgement}
Supported by the Key Research and Development Program of Guangdong Province (grant No. 2021B0101400003) and the corresponding author is Jianzong Wang (\texttt{jzwang@188.com}).

\bibliography{custom}

\begin{thebibliography}{31}
\providecommand{\natexlab}[1]{#1}

\bibitem[{Amrit and Singh(2022)}]{amrit2022survey}
Preetam Amrit and Amit~Kumar Singh. 2022.
\newblock Survey on watermarking methods in the artificial intelligence domain and beyond.
\newblock \emph{Computer Communications}, 188:52--65.

\bibitem[{Ardizzone et~al.(2019)Ardizzone, L{\"u}th, Kruse, Rother, and K{\"o}the}]{ardizzone2019guided}
Lynton Ardizzone, Carsten L{\"u}th, Jakob Kruse, Carsten Rother, and Ullrich K{\"o}the. 2019.
\newblock Guided image generation with conditional invertible neural networks.
\newblock \emph{arXiv preprint arXiv:1907.02392}.

\bibitem[{Behrmann et~al.(2019)Behrmann, Grathwohl, Chen, Duvenaud, and Jacobsen}]{behrmann2019invertible}
Jens Behrmann, Will Grathwohl, Ricky~TQ Chen, David Duvenaud, and J{\"o}rn-Henrik Jacobsen. 2019.
\newblock Invertible residual networks.
\newblock In \emph{International conference on machine learning}, pages 573--582. PMLR.

\bibitem[{Chanchal et~al.(2020)Chanchal, Malathi, and Kumar}]{chanchal2020comprehensive}
M~Chanchal, P~Malathi, and Gireesh Kumar. 2020.
\newblock A comprehensive survey on neural network based image data hiding scheme.
\newblock In \emph{the 4th International Conference on IoT in Social, Mobile, Analytics and Cloud}, pages 1245--1249. IEEE.

\bibitem[{Chen et~al.(2023)Chen, Wu, Liu, Liu, Du, and Wei}]{chen2023wavmark}
Guangyu Chen, Yu~Wu, Shujie Liu, Tao Liu, Xiaoyong Du, and Furu Wei. 2023.
\newblock \href {https://arxiv.org/abs/2308.12770} {Wavmark: Watermarking for audio generation}.
\newblock \emph{CoRR}, abs/2308.12770.

\bibitem[{Defferrard et~al.(2017)Defferrard, Benzi, Vandergheynst, and Bresson}]{defferrard2017fma}
Micha{\"{e}}l Defferrard, Kirell Benzi, Pierre Vandergheynst, and Xavier Bresson. 2017.
\newblock {FMA:} {A} dataset for music analysis.
\newblock In \emph{The 18th International Society for Music Information Retrieval Conference, {ISMIR}}, pages 316--323.

\bibitem[{Dinh et~al.(2015)Dinh, Krueger, and Bengio}]{dinh2015nice}
Laurent Dinh, David Krueger, and Yoshua Bengio. 2015.
\newblock {NICE:} non-linear independent components estimation.
\newblock In \emph{The 3rd International Conference on Learning Representations, {ICLR}, Workshop Track Proceedings}.

\bibitem[{Dinh et~al.(2017)Dinh, Sohl{-}Dickstein, and Bengio}]{dinh2017density}
Laurent Dinh, Jascha Sohl{-}Dickstein, and Samy Bengio. 2017.
\newblock Density estimation using real {NVP}.
\newblock In \emph{The Fifth International Conference on Learning Representations, {ICLR}, Conference Track Proceedings}.

\bibitem[{Goodfellow et~al.(2020)Goodfellow, Pouget-Abadie, Mirza, Xu, Warde-Farley, Ozair, Courville, and Bengio}]{goodfellow2020generative}
Ian Goodfellow, Jean Pouget-Abadie, Mehdi Mirza, Bing Xu, David Warde-Farley, Sherjil Ozair, Aaron Courville, and Yoshua Bengio. 2020.
\newblock Generative adversarial networks.
\newblock \emph{Communications of the ACM}, 63(11):139--144.

\bibitem[{Huang et~al.(2017)Huang, Liu, Van Der~Maaten, and Weinberger}]{huang2017densely}
Gao Huang, Zhuang Liu, Laurens Van Der~Maaten, and Kilian~Q Weinberger. 2017.
\newblock Densely connected convolutional networks.
\newblock In \emph{Conference on Computer Vision and Pattern Recognition}, pages 4700--4708.

\bibitem[{Hussain et~al.(2020)Hussain, Zeng, Qin, and Tan}]{hussain2020survey}
Israr Hussain, Jishen Zeng, Xinhong Qin, and Shunquan Tan. 2020.
\newblock A survey on deep convolutional neural networks for image steganography and steganalysis.
\newblock \emph{KSII Transactions on Internet and Information Systems}, 14(3):1228--1248.

\bibitem[{Kingma and Ba(2015)}]{kingma2015adam}
Diederik~P. Kingma and Jimmy Ba. 2015.
\newblock Adam: {A} method for stochastic optimization.
\newblock In \emph{The 3rd International Conference on Learning Representations, {ICLR}, Conference Track Proceedings}.

\bibitem[{Kingma and Dhariwal(2018)}]{kingma2018glow}
Durk~P Kingma and Prafulla Dhariwal. 2018.
\newblock Glow: Generative flow with invertible 1x1 convolutions.
\newblock \emph{Advances in Neural Information Processing Systems}, 31.

\bibitem[{Lee et~al.(2019)Lee, Lee, Jung, and Kim}]{lee2019surface}
Hyungeol Lee, Eunsil Lee, Jiye Jung, and Junsuk Kim. 2019.
\newblock Surface stickiness perception by auditory, tactile, and visual cues.
\newblock \emph{Frontiers in Psychology}, 10:2135.

\bibitem[{Liu et~al.(2023)Liu, Zhang, Fang, Ma, Zhang, and Yu}]{liu2023dear}
Chang Liu, Jie Zhang, Han Fang, Zehua Ma, Weiming Zhang, and Nenghai Yu. 2023.
\newblock Dear: {A} deep-learning-based audio re-recording resilient watermarking.
\newblock In \emph{The 37th {AAAI} Conference on Artificial Intelligence}, pages 13201--13209.

\bibitem[{Liu et~al.(2019)Liu, Guo, Zhang, Zhu, and Xie}]{liu2019a}
Yang Liu, Mengxi Guo, Jian Zhang, Yuesheng Zhu, and Xiaodong Xie. 2019.
\newblock A novel two-stage separable deep learning framework for practical blind watermarking.
\newblock In \emph{The 27th {ACM} International Conference on Multimedia}, pages 1509--1517.

\bibitem[{Lu et~al.(2021)Lu, Wang, Zhong, and Rosin}]{lu2021large}
Shao-Ping Lu, Rong Wang, Tao Zhong, and Paul~L Rosin. 2021.
\newblock Large-capacity image steganography based on invertible neural networks.
\newblock In \emph{Proceedings of the IEEE/CVF conference on computer vision and pattern recognition}, pages 10816--10825.

\bibitem[{Lugmayr et~al.(2020)Lugmayr, Danelljan, Gool, and Timofte}]{lugmayr2020srflow}
Andreas Lugmayr, Martin Danelljan, Luc~Van Gool, and Radu Timofte. 2020.
\newblock Srflow: Learning the super-resolution space with normalizing flow.
\newblock In \emph{Computer Vision - {ECCV} - The 16th European Conference}, volume 12350, pages 715--732.

\bibitem[{Ma et~al.(2022)Ma, Guo, Hou, Yang, Li, Jia, and Xie}]{ma2022toward}
Rui Ma, Mengxi Guo, Yi~Hou, Fan Yang, Yuan Li, Huizhu Jia, and Xiaodong Xie. 2022.
\newblock Towards blind watermarking: Combining invertible and non-invertible mechanisms.
\newblock In \emph{The 30th {ACM} International Conference on Multimedia}, pages 1532--1542.

\bibitem[{Mou et~al.(2023)Mou, Xu, Song, Zhao, Ghanem, and Zhang}]{mou2023large}
Chong Mou, Youmin Xu, Jiechong Song, Chen Zhao, Bernard Ghanem, and Jian Zhang. 2023.
\newblock Large-capacity and flexible video steganography via invertible neural network.
\newblock In \emph{Proceedings of the IEEE/CVF Conference on Computer Vision and Pattern Recognition}, pages 22606--22615.

\bibitem[{Mushgil et~al.(2018)Mushgil, Adnan, Al-Hadad, and Ahmad}]{mushgil2018efficient}
Baydaa~Mohammad Mushgil, Wan Azizun~Wan Adnan, Syed Abdul-Rahman Al-Hadad, and Sharifah Mumtazah~Syed Ahmad. 2018.
\newblock An efficient selective method for audio watermarking against de-synchronization attacks.
\newblock \emph{Journal of Electrical Engineering and Technology}, 13(1):476--484.

\bibitem[{Pavlović et~al.(2020)Pavlović, Kovačević, and Ðurović}]{pavlovic2020speech}
Kosta Pavlović, Slavko Kovačević, and Igor Ðurović. 2020.
\newblock Speech watermarking using deep neural networks.
\newblock In \emph{The 28th Telecommunications Forum}, pages 1--4.

\bibitem[{Pavlović et~al.(2022)Pavlović, Kovačević, Ðurović, and Wojciechowski}]{pavlovic2022robust}
Kosta Pavlović, Slavko Kovačević, Igor Ðurović, and Adam Wojciechowski. 2022.
\newblock Robust speech watermarking by a jointly trained embedder and detector using a {DNN}.
\newblock \emph{Digital Signal Processing}, 122:103381.

\bibitem[{Prabha and Sam(2022)}]{prabha2022a}
K.~Prabha and I.~Shatheesh Sam. 2022.
\newblock A survey of digital image watermarking techniques in spatial, transform, and hybrid domains.
\newblock \emph{International Journal of Software Innovation}, 10(1):1--21.

\bibitem[{Singh and Singh(2023)}]{singh2023comprehensive}
Himanshu~Kumar Singh and Amit~Kumar Singh. 2023.
\newblock Comprehensive review of watermarking techniques in deep-learning environments.
\newblock \emph{Journal of Electronic Imaging}, 32(3):031804--031804.

\bibitem[{Singh et~al.(2023)Singh, Saraswat, Ashok, Mittal, Tripathi, Pandey, and Pal}]{singh2023classical}
Roop Singh, Mukesh Saraswat, Alaknanda Ashok, Himanshu Mittal, Ashish Tripathi, Avinash~Chandra Pandey, and Raju Pal. 2023.
\newblock From classical to soft computing based watermarking techniques: A comprehensive review.
\newblock \emph{Future Generation Computer Systems}, 141:738--754.

\bibitem[{Singha and Ullah(2022)}]{singha2022development}
Amita Singha and Muhammad~Ahsan Ullah. 2022.
\newblock Development of an audio watermarking with decentralization of the watermarks.
\newblock \emph{Journal of King Saud University-Computer and Information Sciences}, 34(6):3055--3061.

\bibitem[{van~der Ouderaa and Worrall(2019)}]{ouderaa2019reversible}
Tycho F.~A. van~der Ouderaa and Daniel~E. Worrall. 2019.
\newblock Reversible gans for memory-efficient image-to-image translation.
\newblock In \emph{{IEEE} Conference on Computer Vision and Pattern Recognition}, pages 4720--4728.

\bibitem[{Wang et~al.(2020)Wang, Xiao, Liu, Zheng, and Liu}]{wang2020modeling}
Yaolong Wang, Mingqing Xiao, Chang Liu, Shuxin Zheng, and Tie{-}Yan Liu. 2020.
\newblock \href {https://arxiv.org/abs/2006.11999} {Modeling lost information in lossy image compression}.
\newblock \emph{CoRR}, abs/2006.11999.

\bibitem[{Yamagishi et~al.(2016)Yamagishi, Veaux, and MacDonald}]{vctk2016}
Junichi Yamagishi, Christophe Veaux, and Kirsten MacDonald. 2016.
\newblock Cstr vctk corpus: English multi-speaker corpus for cstr voice cloning toolkit.

\bibitem[{Zhang et~al.(2023)Zhang, Zheng, Su, Zeng, and Wang}]{zhang2023m}
Guofu Zhang, Lulu Zheng, Zhaopin Su, Yifei Zeng, and Guoquan Wang. 2023.
\newblock M-sequences and sliding window based audio watermarking robust against large-scale cropping attacks.
\newblock \emph{IEEE Transactions on Information Forensics and Security}, 18:1182--1195.

\end{thebibliography}

\clearpage

\onecolumn
\appendix

\section{Appendix}
\subsection{Training Pipeline of IDEAW}
\label{appendix:alg}
Alg. \ref{alg:algorithm} shows the training pipeline and the acquisition of each loss function. The length regulators draw mappings between the bit sequence space (\textit{i.e.} locating code and message) and a space whose elements are of equal length to the audio waveform segment. Note that $\boldsymbol{x,m,c}$ in the pseudo script refers exclusively to the host audio waveform, message bit sequence and locating code, while the prime $'$ denotes the data in the STFT domain. 

\begin{algorithm*}[!ht]
\caption{Acquisition of total loss $\mathcal{L}_{total}$ in the training stage.}
\label{alg:algorithm}
\begin{flushleft}
\textbf{Input}: \\
\qquad host audio segment $\boldsymbol{x}$, watermark message $\boldsymbol{m}$, locating code $\boldsymbol{c}$ \\
\textbf{Module}: \\
\qquad message embedder $\boldsymbol{Emb_1}$, locating code embedder $\boldsymbol{Emb_2}$, \\
\qquad message extractor $\boldsymbol{Ext_1}$, locating code extractor $\boldsymbol{Ext_2}$, \\
\qquad length regulator $\boldsymbol{LR_1}$, $\boldsymbol{LR_2}$, $\boldsymbol{LR_3}$, $\boldsymbol{LR_4}$, attack layer $\boldsymbol{Att}$, balance block $\boldsymbol{B}$\\
\textbf{Operation}: \\
\qquad short-time Fourier transform $STFT(\cdot)$, inverse short-time Fourier transform $ISTFT(\cdot)$\\
\textbf{Parameter}: \\
\qquad robustness training flag $Robust$, loss weights $\lambda_1,\lambda_2,\lambda_3$ \\
\textbf{Output}: \\
\qquad total loss $\mathcal{L}_{total}$ \\
\end{flushleft}
\begin{algorithmic}[1]
\STATE Regular the length, transform to STFT domain: \\ 
\qquad $\boldsymbol{m'} \leftarrow STFT(\boldsymbol{LR_1}(\boldsymbol{m}))$ \\
\qquad $\boldsymbol{c'} \leftarrow STFT(\boldsymbol{LR_2}(\boldsymbol{c}))$ \\
\qquad $\boldsymbol{x'} \leftarrow STFT(\boldsymbol{x})$ \\
\STATE Embed message: \\
\qquad $\boldsymbol{x'_{w1}} \leftarrow \boldsymbol{Emb_1}(\boldsymbol{x'}, \boldsymbol{m'})$
\STATE First extraction:\\
\qquad $\boldsymbol{\hat{m}_1} \leftarrow \boldsymbol{LR_3}(ISTFT(\boldsymbol{Ext_1}(\boldsymbol{x'_{w1}})))$
\STATE Embed locating code: \\ 
\qquad $\boldsymbol{x'_{w}} \leftarrow \boldsymbol{Emb_2}(\boldsymbol{x'_{w1}}, \boldsymbol{c'})$
\STATE Obtain watermarked audio waveform: \\ 
\qquad $\boldsymbol{x_{w}} \leftarrow ISTFT(\boldsymbol{x'_{w}})$
\STATE Train $\boldsymbol{D}$: \\
\qquad Obtain $\mathcal{L}_{D}$ from $\{\boldsymbol{x}, \boldsymbol{x_w}\}$ \\
\qquad Perform the backward propagation of $\boldsymbol{D}$
\IF{$\boldsymbol{Robust}=True$}
\STATE \quad $\boldsymbol{x_{w}} \leftarrow \boldsymbol{B}(\boldsymbol{Att}(\boldsymbol{x_{w}}))$
\ENDIF
\STATE Extract locating code: \\
\qquad $\boldsymbol{x'}_{mid}, \boldsymbol{\hat{c}'} \leftarrow \boldsymbol{Ext_2}(\boldsymbol{x'_{w}})$ \\
\qquad $\boldsymbol{\hat{c}} \leftarrow \boldsymbol{LR_4}(ISTFT(\boldsymbol{\hat{c}'}))$
\STATE Extract message: \\
\qquad $\boldsymbol{\hat{m}} \leftarrow \boldsymbol{LR_3}(ISTFT(\boldsymbol{\boldsymbol{Ext_1}(\boldsymbol{x'}_{mid})}))$ 
\STATE Obtain losses: \\
\qquad Obtain integrity loss $\mathcal{L}_{integ}$ from $\{\boldsymbol{m},\boldsymbol{\hat{m}_1},\boldsymbol{\hat{m}},\boldsymbol{c},\boldsymbol{\hat{c}}\}$ \\
\qquad Obtain perceptual loss $\mathcal{L}_{percpt}$ from $\{\boldsymbol{x'},\boldsymbol{x'_w}\}$ \\
\qquad Obtain identify loss $\mathcal{L}_{ident}$ from $\{\boldsymbol{x},\boldsymbol{x_{w}}\}$ \\
\qquad $\mathcal{L}_{total} \leftarrow \lambda_1 \mathcal{L}_{integ}+\lambda_2\mathcal{L}_{percept}+\lambda_3\mathcal{L}_{ident}$ \\
\STATE \textbf{return} $\mathcal{L}_{total}$
\end{algorithmic}
\end{algorithm*}

\newpage
\subsection{Configuration of the Attacks}
\label{appendix:attack}
The configuration follows previous works. The descriptions and settings are shown in Table \ref{tab:attack}. 

\begin{table*}[htb]
    \centering
    	\begin{tabularx}{\textwidth}{llX}
		\toprule
		ID & Attack & Description and Configuration  \\
		\midrule
		GN & Gaussian additive noise
		& Add Gaussian noise to the watermarked audio and maintain the signal-to-noise ratio at approximately $35 dB$. 
		\\
        LF & lower-pass filter 
        &  Pass the audio through a lower-pass filter of $5 kHz$, the range of the filter is set according to the human auditory range
        \\
        CP & MP3 compression 
        & Compress the waveform to $64 kbps$ \textit{MP3} format and then convert back to \textit{wav} format. This conversion process results in information loss.
        \\
        QZ & quantization 
        &  Quantize the sample points of watermarked audio waveform to $2^9$ levels. 
        \\
        RD & random dropout 
        &  Randomly value $0.1\%$ of the sample points in the watermarked audio to zero.
        \\
        RS & resampling 
        &  Resample the audio to a new sampling rate ($200\%$ of the original sample rate) then resample back to the original sample rate.
        \\
        AM & amplitude modification 
        & Multiply $90\%$ to the overall amplitude of the audio by a modification factor.
        \\
        TS & time stretch 
        & Compress the audio in the time domain to $90\%$ of the original length, then stretch it to maintain the original length.
        \\
		\bottomrule
	\end{tabularx}%
    \caption{Descriptions and settings of the attacks.}
    \label{tab:attack}
\end{table*}

\end{document}